# Designed Sampling from Large Databases for Controlled Trials

Liwen Ouyang[a], Daniel W. Apley[a], Sanjay Mehrotra[a]
[a]Department of Industrial Engineering and Management Sciences,
Northwestern University,
Evanston, IL 60208

**Corresponding author**:
Daniel W. Apley
2145 Sheridan Road
Tech C150
Evanston, IL 60208-3109
Phone: (847) 491-2397
Email: apley@northwestern.edu

**Abstract**: The increasing prevalence of rich sources of data and the availability of electronic medical record databases and electronic registries opens tremendous opportunities for enhancing medical research. For example, controlled trials are ubiquitously used to investigate the effect of a medical treatment, perhaps dependent on a set of patient covariates, and traditional approaches have relied primarily on randomized patient sampling and allocation to treatment and control group. However, when covariate data for a large cohort group of patients have already been collected and are available in a database, one can potentially design a treatment/control sample and allocation that provides far better estimates of the covariate-dependent effects of the treatment. In this paper, we develop a new approach that uses optimal design of experiments (DOE) concepts to accomplish this objective. The approach selects the patients for the treatment and control samples upfront, based on their covariate values, in a manner that optimizes the information content in the data. For the optimal sample selection, we develop simple guidelines and an optimization algorithm that provides solutions that are substantially better than random sampling. Moreover, our approach causes no sampling bias in the estimated effects, for the same reason that DOE principles do not bias estimated effects. We test our method with a simulation study based on a testbed data set containing information on the effect of statins on low-density lipoprotein (LDL) cholesterol.

**Keywords:** controlled trials, randomized trials, design of experiments, sampling from databases,

# 1 INTRODUCTION

As more and higher-variety data become increasingly available in the information age, this offers revolutionary opportunities for enhancing medical research, e.g., in evaluating the effects of medical treatments in situation in which the effects depend on a set of patient covariates such as patient age, gender, ethnicity, etc. Controlled trials are used to investigate the effect of a treatment, and traditional controlled trials typically adopt random sampling and allocation of patients to the treatment and control groups in order to avoid biasing the treatment assignment and the estimated treatment effects [1]. These so-called randomized controlled trials involve several different types of randomization approaches, the simplest of which is unrestricted randomization. Also frequently used are restricted randomization approaches, including blocked randomization and stratified randomization [1-3]. A third type of approach is adaptive randomization, including covariate-adaptive randomization and response-adaptive randomization [2-4]. Most previous literature of the latter type focuses on the situation in which patients arrive sequentially, and, as each new patient arrives, one must decide to which group (treatment or control) the new patient should be allocated. There is also a large body of work on propensity scoring [5], in which one models the probability that a patient is selected for treatment as a function of the covariates and then incorporates this propensity score model into the analysis. However, propensity scoring is largely inapplicable to our problem, as it applies to observational data in which patients have already been selected for treatment based on other considerations.

In contrast, taking advantage of the availability of databases containing covariate information on large numbers of patients, our proposed approach is to select upfront (in advance of conducting the study) a small but informative subset of patients from the database and to allocate each of the selected patients to either the treatment or the control group. The goal is to select and allocate the patients judiciously, so as to maximize the amount of information in the data (as it pertains to estimating the covariate-dependent effects of the treatment factor) for a specified sample size. To motivate and illustrate the approach, consider a health management provider that contracts with large companies to offer diabetes outreach (and many other health management services) to employees of the companies. They would like to conduct a study to assess whether participation by an at-risk patient in a new diabetes outreach program (the treatment factor) is effective at lowering the patient's risk (as measured by an appropriate response variable); and, if so, they would like to quantify the effect of the treatment, recognizing that the effect may depend on the levels of a number of other covariates, such as various demographic, blood test and other health-related measures for that patient. They have such covariate information available for a large number of patients in their database, based on prior participation by the patients in other health management activities. Some form of controlled trial must be used, in which a small fraction of patients in the database are selected and allocated to either the control group or the treatment group.

In this paper, we consider the more general situation in which the goal is to select a sample of patients, for participation in a controlled trial study, from an existing database containing patient covariates and then to allocate each patient to either the control or treatment group. Rather than using random sampling and allocation, we propose an optimal design of experiments (DOE) approach, in which patients are judiciously selected from the database and allocated based on their covariates. The DOE goal is to select and allocate the sample to maximize the information content of the data, as it pertains to estimating the covariate-dependent effects of the treatment factor. This is a computationally intractable combinatorial optimization problem using today's technology. In light of this, we develop simple guidelines that provide insight into the nature of the optimal sample and, based on this, an optimization algorithm for selecting the optimal sample. We refer to the approach as designed sampling from databases (DSD) for controlled trials, and we demonstrate that it provides a far more informative study than random sampling and allocation.

We note that previous literature on covariate-adaptive randomization methods [6-10] has considered DOE-based criteria to allocate patients to treatment and control groups. However, the prior work considers patients that arrive sequentially, and each patient that arrives is allocated to the treatment or control group using DOE criteria. In contrast, in the present work we are also using DOE concepts to select the most appropriate sample of patients for the study, from among a very large set of patients in an existing database. Moreover, we also use a batch DOE strategy that selects and allocates the entire sample of patients in one shot, as opposed to a sequential DOE strategy that allocates patients one-at-a-time. For a specified sample size, optimal batch designs are generally more powerful than their suboptimal sequential design counterparts.

It should be emphasized that although our DSD approach does not use random sampling, it introduces no sampling bias that might bias the estimated coefficients of the model. On first thought, one might speculate that any non-random sampling strategy may introduce a sampling bias. However, our DSD approach selects the sample based on patients' covariate data, without consideration of their response data. This is precisely what happens in any DOE, when one selects the factor settings (analogous to our selected patient covariate values) for each experimental run (analogous to one selected patient in the sample), without knowledge of the yet-to-be-realized response value for that run. Consequently, for the same reason that DOE does not result in a biased estimated model, neither does our DSD approach. Moreover, for the same reason that designed experiments with carefully selected factor settings generally result in far more informative data than randomly selected factor settings, our DSD approach results in more informative data than random patient selection.

The format of the remainder of the paper is as follows. In Section 2, we overview the DSD problem and describe the DSD sample design concepts. In Section 3, we describe the optimization algorithm for selecting and allocating the optimal DSD sample. In Section 4, we illustrate and evaluate the approach with an example. Section 5 concludes the paper and briefly discusses issues such as nonrandom sampling and bias, missing data, and the effect of patient nonparticipation.

## 2 DSD SAMPLE DESIGN PRINCIPLES

### 2.1 Problem overview

We now formally describe the general underlying statistical model considered in this paper. Assume we have available a database containing covariate information (e.g., demographic and phenotypic information) for a large cohort group of $N$ patients. We are interested in the effect of a particular treatment on a response variable $Y$ for the patients. Let $z = \pm 1$ denote the treatment variable for a patient, i.e., $z = -1$ if the patient is in the control group, and $z = 1$ if the patient is in the treatment group. Suppose we intend to design and collect a sample of size $2n$ patients from this large cohort group of $N$ patients ($2n \ll N$ typically), with $n$ patients allocated to the control group and $n$ allocated to the treatment group. Also suppose we have $k$ covariates $\{x_1, x_2, \ldots, x_k\}$ available for each patient, and the treatment effect may depend on the covariates. To represent this situation, we assume the standard linear model with interactions between treatment and covariates:

$$Y_i = \delta + \alpha z_i + \beta_1 z_i x_{i1} + \ldots + \beta_k z_i x_{ik} + \gamma_1 x_{i1} + \ldots + \gamma_k x_{ik} + \varepsilon_i, \quad i = 1, 2, \ldots, 2n \qquad (1)$$

where the subscript $i$ is the patient index (or row index in the data array to which the regression model will be fit), $\varepsilon_i$ is the random error term for the $i$th patient, $\boldsymbol{\theta} = [\delta \ \alpha \ \boldsymbol{\beta}^T \ \boldsymbol{\gamma}^T]^T$ are the model parameters to be estimated, $\boldsymbol{\beta} = [\beta_1, \beta_2, \ldots, \beta_k]^T$, and $\boldsymbol{\gamma} = [\gamma_1, \gamma_2, \ldots, \gamma_k]^T$.

Rewriting the model (1) as (and omitting the patient index)

$$Y = \delta + (\alpha + \beta_1 x_1 + \ldots + \beta_k x_k)z + \gamma_1 x_1 + \ldots + \gamma_k x_k, \tag{2}$$

it follows that for a fixed set of covariate values $\{x_1, x_2, \ldots, x_k\}$ for a patient, the covariate-dependent effect of the treatment is $\alpha + \beta_1 x_1 + \ldots + \beta_k x_k$. Consequently, our primary interest is in the parameters $\alpha$ and $\boldsymbol{\beta}$, and our objective is to select and allocate the sample of $2n$ patients so as to achieve the best possible estimates of these parameters of interest. We briefly discuss the issue of patient nonparticipation and missing covariate data in Section 5.

**2.2 Sample design concepts**

We use DOE principles to select and allocate the most informative patients to comprise the sample of $2n$ patients, from the database of $N$ patients, for the controlled trial study. In the traditional DOE literature, a number of methods have been developed to select the values of the design variables $\mathbf{x} = \{x_1, x_2, \ldots, x_k\}$ for the experimental cases so as to optimize some measure of quality of the resulting fitted model. The most common optimal DOE criteria when fitting linear regression models are the so-called "alphabetic" optimality criteria (e.g., D-optimality) [11,12], and algorithmic design optimization is quite well developed [13-15].

Existing optimal DOE methods are not directly applicable to our situation, because of our restriction that we must sample patients from some larger set of patients in an existing database. First, our design space (the set of feasible values for $\mathbf{x}$) is neither a continuous domain (common in traditional DOE when the input factors are continuous variables), nor a region or subregion of a grid (common in traditional DOE when the input factors have discrete settings). Our design space is the set of existing **x** values for the *N* patients in the database. For large *N*, there are a great many possible **x** values to consider, and these data points typically have quite irregular structure due to issues such as multicollinearity, clusters, outliers, etc. in the **x**-space. Second, when considering whether/how to modify an **x** value in the design, its elements cannot be modified independently, as in the coordinate exchange algorithm of Meyer and Nachtsheim [15] (the most widely used DOE optimization algorithm), because only the fixed set of **x** values in the available data set are permissible. Our DOE optimization approach is closest to that of Ouyang, et al. [16], who developed an optimal DOE-based algorithm for selecting a validation sample of medical records for chart review from a large database of unreliable, error-prone records. Their optimal DOE criterion pertained to fitting a logistic regression model to the validated data, but they did not consider the problem of allocating patients to control/treatment groups in a controlled trial study.

In this paper, we use a D-optimality criterion to select and allocate our sample of $2n$ patients. That is, we choose the sample to maximize the determinant of Fisher information matrix for the fitted linear regression model. To develop this approach, rewrite Model (1) in regression matrix form as

$$\mathbf{Y} = \mathbf{M}\boldsymbol{\theta} + \boldsymbol{\varepsilon}, \tag{3}$$

where $\mathbf{Y} = [Y_1, Y_2, \ldots, Y_{2n}]^T$, and $\boldsymbol{\varepsilon} = [\varepsilon_1, \varepsilon_2, \ldots, \varepsilon_{2n}]^T$ with each $\varepsilon_i \sim N(0, \sigma^2)$ assumed Gaussian and independent. It is straightforward to show that the overall regression design matrix is

$$\mathbf{M} = \begin{bmatrix} \mathbf{1} & -\mathbf{1} & -\mathbf{X}_- & \mathbf{X}_- \\ \mathbf{1} & \mathbf{1} & \mathbf{X}_+ & \mathbf{X}_+ \end{bmatrix},$$

where **1** is an *n*×1 vector of ones, $\mathbf{X}_-$ is the *n*×*k* design matrix for the group of *n* control patients, and $\mathbf{X}_+$ is the *n*×*k* design matrix for the group of *n* treatment patients. That is,

$$\mathbf{X}_- = \begin{bmatrix} x_{11} & \cdots & x_{1k} \\ \vdots & & \vdots \\ x_{n1} & \cdots & x_{nk} \end{bmatrix}$$

for the group of $n$ control patients, with the first subscript on $x_{ij}$ indicating the patient index. The $n \times k$ matrix $\mathbf{X}_+$ is defined similarly, but for the group of $n$ treatment patients.

From standard linear regression theory [17], the Fisher information matrix is $\mathbf{F} = \frac{1}{\sigma^2}\mathbf{M}^T\mathbf{M}$, and the covariance matrix of the estimated regression parameter $\widehat{\boldsymbol{\theta}}$ is $\mathbf{F}^{-1}$. In our case,

$$\mathbf{F} = \frac{1}{\sigma^2} \begin{bmatrix} \mathbf{1}^T & \mathbf{1}^T \\ -\mathbf{1}^T & \mathbf{1}^T \\ -\mathbf{X}_-^T & \mathbf{X}_+^T \\ \mathbf{X}_-^T & \mathbf{X}_+^T \end{bmatrix} \begin{bmatrix} \mathbf{1} & -\mathbf{1} & -\mathbf{X}_- & \mathbf{X}_- \\ \mathbf{1} & \mathbf{1} & \mathbf{X}_+ & \mathbf{X}_+ \end{bmatrix}$$

$$= \frac{n}{\sigma^2} \begin{bmatrix} 2 & 0 & \bar{\mathbf{x}}_+^T - \bar{\mathbf{x}}_-^T & \bar{\mathbf{x}}_+^T + \bar{\mathbf{x}}_-^T \\ 0 & 2 & \bar{\mathbf{x}}_+^T + \bar{\mathbf{x}}_-^T & \bar{\mathbf{x}}_+^T - \bar{\mathbf{x}}_-^T \\ \bar{\mathbf{x}}_+^T - \bar{\mathbf{x}}_-^T & \bar{\mathbf{x}}_+^T + \bar{\mathbf{x}}_-^T & \mathbf{S}_+ + \mathbf{S}_- & \mathbf{S}_+ - \mathbf{S}_- \\ \bar{\mathbf{x}}_+^T + \bar{\mathbf{x}}_-^T & \bar{\mathbf{x}}_+^T - \bar{\mathbf{x}}_-^T & \mathbf{S}_+ - \mathbf{S}_- & \mathbf{S}_+ + \mathbf{S}_- \end{bmatrix}, \quad (4)$$

where $\mathbf{S}_+ = \frac{1}{n}\mathbf{X}_+^T\mathbf{X}_+$ and $\mathbf{S}_- = \frac{1}{n}\mathbf{X}_-^T\mathbf{X}_-$ are similar to the $k \times k$ sample covariance matrices (except that the sample average is not subtracted from the data) for the vector $\mathbf{x} = \{x_1, x_2, \ldots, x_k\}$ over the treatment and control groups, respectively, and $\bar{\mathbf{x}}_+$ and $\bar{\mathbf{x}}_-$ are the $k \times 1$ sample average vectors for $\mathbf{x}$ over the treatment and control groups, respectively.

From (4), we can reduce aliasing between the estimates of $\alpha$, $\boldsymbol{\beta}$, and $\boldsymbol{\gamma}$ by imposing the following design constraints (which will be ensured later in the optimization algorithm):

$$\bar{\mathbf{x}}_+ \approx \bar{\mathbf{x}}_- \quad \text{and} \quad \mathbf{S}_+ \approx \mathbf{S}_- \quad (5)$$

The interpretation of these constraints is that we require that the sample average vectors and sample covariance matrices of $\mathbf{x}$ to be similar for the control and treatment groups. This clearly has intuitive appeal, as one would expect that the patients chosen for the treatment group should be statistically similar to the patients chosen for the control group. It also has important implications that enable a computationally feasible, two-stage design optimization algorithm (see Section 3). With these constraints, the Fisher information matrix (4) becomes

$$\mathbf{F} \cong \frac{n}{\sigma^2} \begin{bmatrix} 2 & 0 & \mathbf{0} & \bar{\mathbf{x}}_+^T + \bar{\mathbf{x}}_-^T \\ 0 & 2 & \bar{\mathbf{x}}_+^T + \bar{\mathbf{x}}_-^T & \mathbf{0} \\ \mathbf{0} & \bar{\mathbf{x}}_+ + \bar{\mathbf{x}}_- & \mathbf{S}_+ + \mathbf{S}_- & \mathbf{0} \\ \bar{\mathbf{x}}_+ + \bar{\mathbf{x}}_- & \mathbf{0} & \mathbf{0} & \mathbf{S}_+ + \mathbf{S}_- \end{bmatrix}$$

$$= \frac{n}{\sigma^2} \begin{bmatrix} 2 & 0 & \mathbf{0} & 2\bar{\mathbf{x}}^T \\ 0 & 2 & 2\bar{\mathbf{x}}^T & \mathbf{0} \\ \mathbf{0} & 2\bar{\mathbf{x}} & \frac{1}{n}\mathbf{X}^T \mathbf{X} & \mathbf{0} \\ 2\bar{\mathbf{x}} & \mathbf{0} & \mathbf{0} & \frac{1}{n}\mathbf{X}^T \mathbf{X} \end{bmatrix}, \qquad (6)$$

where $\mathbf{0}$ denotes a vector or matrix of zeros of appropriate dimension, $\mathbf{X} = \begin{bmatrix} \mathbf{X}_- \\ \mathbf{X}_+ \end{bmatrix}$ is the $2n \times k$ design matrix for $\mathbf{x}$, and $\bar{\mathbf{x}}$ is $k \times 1$ average vector for $\mathbf{x}$, both of which are over the entire sample of $2n$ patients.

By inverting the expression (6) and using results on the inverse of a partitioned matrix [18], the covariance matrix of $\hat{\boldsymbol{\theta}}$ is

$$\mathbf{F}^{-1} \cong \frac{\sigma^2}{2n} \begin{bmatrix} \frac{1}{1 - 2n\bar{\mathbf{x}}^T(\mathbf{X}^T\mathbf{X})^{-1}\bar{\mathbf{x}}} & 0 & \mathbf{0} & -\bar{\mathbf{x}}^T\left(\frac{\mathbf{X}^T\mathbf{X}}{2n} - \bar{\mathbf{x}}\bar{\mathbf{x}}^T\right)^{-1} \\ 0 & \frac{1}{1 - 2n\bar{\mathbf{x}}^T(\mathbf{X}^T\mathbf{X})^{-1}\bar{\mathbf{x}}} & -\bar{\mathbf{x}}^T\left(\frac{\mathbf{X}^T\mathbf{X}}{2n} - \bar{\mathbf{x}}\bar{\mathbf{x}}^T\right)^{-1} & \mathbf{0} \\ \mathbf{0} & -\left(\frac{\mathbf{X}^T\mathbf{X}}{2n} - \bar{\mathbf{x}}\bar{\mathbf{x}}^T\right)^{-1}\bar{\mathbf{x}} & \left(\frac{\mathbf{X}^T\mathbf{X}}{2n} - \bar{\mathbf{x}}\bar{\mathbf{x}}^T\right)^{-1} & \mathbf{0} \\ -\left(\frac{\mathbf{X}^T\mathbf{X}}{2n} - \bar{\mathbf{x}}\bar{\mathbf{x}}^T\right)^{-1}\bar{\mathbf{x}} & \mathbf{0} & \mathbf{0} & \left(\frac{\mathbf{X}^T\mathbf{X}}{2n} - \bar{\mathbf{x}}\bar{\mathbf{x}}^T\right)^{-1} \end{bmatrix}$$

$$= \frac{\sigma^2}{2n} \begin{bmatrix} 1 + \bar{\mathbf{x}}^T\mathbf{S}^{-1}\bar{\mathbf{x}} & 0 & \mathbf{0} & -\bar{\mathbf{x}}^T\mathbf{S}^{-1} \\ 0 & 1 + \bar{\mathbf{x}}^T\mathbf{S}^{-1}\bar{\mathbf{x}} & -\bar{\mathbf{x}}^T\mathbf{S}^{-1} & \mathbf{0} \\ \mathbf{0} & -\mathbf{S}^{-1}\bar{\mathbf{x}} & \mathbf{S}^{-1} & \mathbf{0} \\ -\mathbf{S}^{-1}\bar{\mathbf{x}} & \mathbf{0} & \mathbf{0} & \mathbf{S}^{-1} \end{bmatrix}, \qquad (7)$$

where $\mathbf{S} = \frac{\mathbf{X}^T\mathbf{X}}{2n} - \bar{\mathbf{x}}\bar{\mathbf{x}}^T$ is the sample covariance matrix of $\mathbf{x}$ over the entire sample of $2n$ patients, and we have used the relationship $(\mathbf{X}^T\mathbf{X})^{-1} = \frac{1}{2n}\left[\mathbf{S}^{-1} - \frac{\mathbf{S}^{-1}\bar{\mathbf{x}}\bar{\mathbf{x}}^T\mathbf{S}^{-1}}{1+\bar{\mathbf{x}}^T\mathbf{S}^{-1}\bar{\mathbf{x}}}\right]$.

From (7), the covariance matrix of the regression estimates of $\alpha$ and $\boldsymbol{\beta}$ is

$$Cov(\hat{\alpha}, \widehat{\boldsymbol{\beta}}) = \frac{\sigma^2}{2n}\begin{bmatrix} 1 + \bar{\mathbf{x}}^T\mathbf{S}^{-1}\bar{\mathbf{x}} & -\bar{\mathbf{x}}^T\mathbf{S}^{-1} \\ -\mathbf{S}^{-1}\bar{\mathbf{x}} & \mathbf{S}^{-1} \end{bmatrix}. \tag{8}$$

Because of the symmetry and block structure of (7), it follows that $Cov(\hat{\delta}, \hat{\boldsymbol{\gamma}}) = Cov(\hat{\alpha}, \widehat{\boldsymbol{\beta}})$, and that $\{\hat{\delta}, \hat{\boldsymbol{\gamma}}\}$ are uncorrelated (unaliased) with $\{\hat{\alpha}, \widehat{\boldsymbol{\beta}}\}$. Consequently, $|\mathbf{F}^{-1}| = |Cov(\hat{\alpha}, \widehat{\boldsymbol{\beta}})| \times |Cov(\hat{\delta}, \hat{\boldsymbol{\gamma}})| = |Cov(\hat{\alpha}, \widehat{\boldsymbol{\beta}})|^2$, so that minimizing the overall D-optimality criterion $|\mathbf{F}^{-1}|$ is equivalent to minimizing $|Cov(\hat{\alpha}, \widehat{\boldsymbol{\beta}})|$, which is the D-optimality criterion for the parameters of primary interest. Using (8) and results on the determinant of a partitioned matrix [19], we have

$$|Cov(\hat{\alpha}, \widehat{\boldsymbol{\beta}})| = \left(\frac{\sigma^2}{2n}\right)^{k+1}|\mathbf{S}^{-1}| \times |1 + \bar{\mathbf{x}}^T\mathbf{S}^{-1}\bar{\mathbf{x}} - \bar{\mathbf{x}}^T\mathbf{S}^{-1}\mathbf{S}\mathbf{S}^{-1}\bar{\mathbf{x}}| = \left(\frac{\sigma^2}{2n}\right)^{k+1}|\mathbf{S}^{-1}|. \tag{9}$$

Consequently, our DSD optimal design strategy is simply to minimize $|\mathbf{S}^{-1}|$ or, equivalently, to maximize $|\mathbf{S}|$, the determinant of the sample covariance matrix of the full sample of $2n$ patients, while ensuring constraints (5). This has important implications regarding the tractability and computational feasibility of the optimization algorithm for selecting and allocating the patients, as discussed in Section 3.

**3 DESIGN OPTIMIZATION ALGORITHM**

Henceforth, we assume that, as a preprocessing step, one has standardized the $k$ covariates by subtracting their average values and dividing by their standard deviations over all $N$ patients in the database. This is necessary for the criteria like D-optimality to be meaningful, and it is also recommended for improving numerical issues in regression analyses.

In light of the developments in Section 2.2, we use a two-stage algorithm to optimize the design. Specifically, in order to minimize $|Cov(\hat{\alpha}, \hat{\boldsymbol{\beta}})|$, Eq. (6c) shows that the entire sample of $2n$ patients can be chosen to maximize the determinant of $\mathbf{S}$, the sample covariance matrix of the covariates for the $2n$ patients. The salient implication of this result is that the $2n$ patients can be chosen without regard to the treatment versus control groups allocation decisions (providing constraints (5) are satisfied), resulting in a far more computationally tractable optimization problem.

Consequently, in Stage 1 of the design algorithm, we select $2n$ patients from the large database of $N$ patients to maximize $|\mathbf{S}| = \left|\frac{\mathbf{X}^T \mathbf{X}}{2n} - \bar{\mathbf{x}}\bar{\mathbf{x}}^T\right|$, which is based entirely on the covariate values of the patients. In Stage 2, we will allocate these $2n$ patients into the treatment or control groups to ensure that $\bar{\mathbf{x}}_+ \approx \bar{\mathbf{x}}_-$ and $\mathbf{S}_+ \approx \mathbf{S}_-$, i.e., to ensure constraints (5) that the sample mean vectors and covariance matrices for the treatment and control groups are equivalent. The remainder of this section describes the optimization algorithms for accomplishing this.

Regarding Stage 1, choosing the $2n$ out of $N$ patients that exactly maximize $\left|\frac{\mathbf{X}^T \mathbf{X}}{2n} - \bar{\mathbf{x}}\bar{\mathbf{x}}^T\right|$ is a computationally intractable combinatorial optimization problem for large $N$. Consequently, we use a backward stepwise (greedy) approach by which we start with all $N$ patients in the sample and, at each step, remove the single patient that least reduces $\left|\frac{\mathbf{X}^T \mathbf{X}}{2n} - \bar{\mathbf{x}}\bar{\mathbf{x}}^T\right|$, stopping when we have reduced the number of patients to the desired $2n$. For a related optimization problem, in terms of solution quality and computational complexity, Ouyang, et al. [16] found the backwards stepwise approach to be more effective than a forwards stepwise approach or a forwards/backwards hybrid approach.

Let $p$ denote the number of patients in the sample at the current step, let $\bar{\mathbf{x}}^p$ denote the $k \times 1$ sample average column vector for the $p$ patients, and let $\mathbf{X}^p$ denote the $p \times k$ design matrix for the $p$ patients. Selecting the next patient to remove from the sample requires calculating $\left|\frac{(\mathbf{X}^{p-1})^T \mathbf{X}^{p-1}}{p-1} - \bar{\mathbf{x}}^{p-1}(\bar{\mathbf{x}}^{p-1})^T\right|$ a total of $p$ times (once for removing each of the $p$ patients), the computational expense of which can be substantially reduced using the following recursive expressions. Let $r$ denote the index of the patient considered for removal and $\mathbf{x}_r$ their $k \times 1$ vector of predictors. Noting that $(\mathbf{X}^{p-1})^T \mathbf{X}^{p-1} = (\mathbf{X}^p)^T \mathbf{X}^p - \mathbf{x}_r(\mathbf{x}_r)^T$, we can recursively (going from $p \to p-1$) update all needed quantities via

$$|(\mathbf{X}^{p-1})^T \mathbf{X}^{p-1}| = |(\mathbf{X}^p)^T \mathbf{X}^p|(1 - (\mathbf{x}_r)^T [(\mathbf{X}^p)^T \mathbf{X}^p]^{-1} \mathbf{x}_r), \tag{10}$$

$$\bar{\mathbf{x}}^{p-1} = \frac{p\bar{\mathbf{x}}^p - \mathbf{x}_r}{p-1},$$

$$[(\mathbf{X}^{p-1})^T \mathbf{X}^{p-1}]^{-1} = [(\mathbf{X}^p)^T \mathbf{X}^p]^{-1} + \frac{[(\mathbf{X}^p)^T \mathbf{X}^p]^{-1} \mathbf{x}_r (\mathbf{x}_r)^T [(\mathbf{X}^p)^T \mathbf{X}^p]^{-1}}{1 - (\mathbf{x}_r)^T [(\mathbf{X}^p)^T \mathbf{X}^p]^{-1} \mathbf{x}_r}, \text{ and} \tag{11}$$

$$\left|\frac{(\mathbf{X}^{p-1})^T \mathbf{X}^{p-1}}{p-1} - \bar{\mathbf{x}}^{p-1}(\bar{\mathbf{x}}^{p-1})^T\right|$$
$$= \left(\frac{1}{p-1}\right)^k |(\mathbf{X}^{p-1})^T \mathbf{X}^{p-1}|(1 - (p-1)(\bar{\mathbf{x}}^{p-1})^T [(\mathbf{X}^{p-1})^T \mathbf{X}^{p-1}]^{-1} \bar{\mathbf{x}}^{p-1}). \tag{12}$$

Eqs. (10)—(12) follow from standard results on the determinant and the inverse of a rank-one modification of a matrix [20,21]. Eqs. (10)—(12) are iteratively executed for $p = N, N-1, \ldots, 2n+1$, until the desired sample size $2n$ is reached. On each iteration, they are repeated for each patient (i.e., each $\mathbf{x}_r$) in the current sample of $p$ patients.

In Stage 2, after the $2n$ patients are selected in Stage 1, half of the $2n$ patients must be allocated to the control group and the other half to the treatment group. The Stage 2 criterion is to ensure that $\bar{\mathbf{x}}_+ \approx \bar{\mathbf{x}}_-$ and $\mathbf{S}_+ \approx \mathbf{S}_-$ per constraints (5), in which case the overall optimization criterion $|Cov(\hat{\alpha}, \hat{\boldsymbol{\beta}})|$ depends only on the overall $\mathbf{S}$ for all $2n$ patients. We begin by randomly allocating $n$ patients to the control group (comprising the rows of $\mathbf{X}_-$) and the other $n$ patients to the treatment group (comprising the rows of $\mathbf{X}_+$), for which the constraints (5) will typically be approximately satisfied. We then follow up by exchanging a pair of patients between the treatment and control groups, with the pair chosen to minimize the criterion $||\bar{\mathbf{x}}_+ - \bar{\mathbf{x}}_-||^2 + ||\mathbf{S}_+ - \mathbf{S}_-||^2$ after the exchange. This exchange procedure is repeated until no further reduction in the criterion can be achieved. Each exchange requires searching over $n^2$ possible pairs of patients.

## 4 EXAMPLE AND PERFORMANCE EVALUATION

To evaluate our approach, we used a testbed data set containing information on the effect of statins (the treatment variable, *z*) on low-density lipoprotein (LDL) cholesterol (the response variable, *y*). The results are not meant to represent scientific findings, but rather to illustrate the capabilities of our DSD method. We considered four covariates ($x_1$ to $x_4$), namely age, body mass index (bmi), diastolic blood pressure (bp_diastolic) and triglycerides (tri). A total of $N = 11,080$ patient records of these four covariates were considered from Northwestern Memorial Hospital EMR database from January 2006 to December 2010. The average and standard deviation are, respectively, 58 and 13.5 (for age), 30.19 and 7.45 (for bmi); 77.6 and 11.76 (bp_diastolic), and 117.23 and 71.43 (tri). All four covariates values were standardized prior to the subsequent analyses.

We applied our DSD approach to select a sample of $2n = 1,000$ patients for a hypothetical controlled-trial study, based on their covariate values for the real data set. To evaluate the performance of the approach, we calculated mean square errors (MSEs) for the parameters of interest that would result from fitting the regression model (1) to the selected sample. Notice that the MSEs can be obtained directly from the diagonal elements of the inverse of **F** in Eq. (4), which depends only on *n*, $\sigma$, and the covariate values for the selected sample. Notice also that $\sigma^2$ multiplies $\mathbf{F}^{-1}$, and, hence, the relative MSE performance when comparing two different methods for selecting the sample is independent of $\sigma$. Consequently, we will assume a single value $\sigma = 0.3$ in the following comparisons, which was approximately the standard deviation of LDL cholesterol over the full data set.

We used Monte Carlo (MC) simulation with 10,000 MC replicates to compare the MSEs of the estimated parameters $\hat{\boldsymbol{\theta}}$ that resulted from using our DSD approach to select and allocate the sample, versus randomly selecting a total $2n$ patients and then randomly allocating half of them to the treatment group and the other half to the control group. For the random sampling method, on each MC replicate we selected a different random sample of $2n$ patients. For our DSD method, the sample of $2n$ patients do not vary from replicate to replicate, since they are selected from the original dataset of $N$ patients. For this example, the MSEs after the final Stage 2 allocation of the DSD method were very consistent for different "initial guess" random allocations at the beginning of Stage 2. More generally, if the MSEs are found to depend on the initial guess, then one should use multiple initial guesses and choose the one that results in the smallest MSEs. The average MSEs (averaged across the 10,000 MC replicates) for the treatment factor ($\hat{\alpha}$) and treatment-by-covariate interactions ($\hat{\boldsymbol{\beta}}$) are shown in Table 1.

Remarkably, Table 1 shows that the MSEs for the estimates of the interaction terms ($\hat{\boldsymbol{\beta}}$) from our DSD approach are one-third to one-fourth the MSEs from random sampling. Notice that the MSE for $\hat{\alpha}$ is very slightly larger (worse) for the DSD method than for random sampling. This is not surprising, as it is straightforward to show from (4) that a lower bound on the MSE of $\hat{\alpha}$ is $\frac{\sigma^2}{2n}$, and this lower bound is achieved when $\bar{\mathbf{x}}_+ = \bar{\mathbf{x}}_- = \mathbf{0}$ is satisfied exactly. Stage 2 of our DSD algorithm ensures that $\bar{\mathbf{x}}_+ \cong \bar{\mathbf{x}}_-$, but not necessarily that they are both close to $\mathbf{0}$ (although the latter often will also be the case, since the covariates have been standardized over the complete set of $N$ patients). When sample size $2n$ is large, pure random sampling typically does a good job of satisfying the condition $\bar{\mathbf{x}}_+ \cong \bar{\mathbf{x}}_- \cong \mathbf{0}$. Hence, it has a slightly lower MSE for $\hat{\alpha}$. However, it is important to recognize that the dramatic reduction in the MSE of $\hat{\boldsymbol{\beta}}$ for the DSD approach, relative to random sampling, far outweighs the slight increase in the MSE of $\hat{\alpha}$.

## 5  CONCLUSIONS

This paper presents a method that uses DOE concepts to judiciously select patients for a controlled trial study, from among a large patient database containing patient covariate information. The two-stage optimization algorithm in our DSD approach first selects in Stage 1 the overall sample of $2n$ patients to enter into the study, and then in Stage 2 subsequently allocates the $2n$ patients in the selected cohort to either treatment and control groups. Interestingly, we have shown that in order to optimally select the patients, it is sufficient to consider the sample covariance matrix of the patient covariates over the entire sample of $2n$ patients, prior to consideration of their allocation to treatment/control groups. Then, in Stage 2, one simply allocates the $2n$ patients to either the treatment group or control group to ensure that the within-group mean vectors and covariance matrices are approximately equal. Exploiting these findings, we have developed a powerful, yet computationally tractable, optimization procedure for optimal sample design. We have demonstrated via an example that the proposed DSD approach can result in a substantially lower MSE for the estimates of the covariate-dependent effects of the treatment, relative to random sampling. From an alternative viewpoint, our findings imply that to ensure a desired MSE for the parameter estimates, much lower sample sizes are needed when using our DSD approach to select patients from a database, versus random sampling. This would result in significant cost savings when conducting such studies.

We reiterate that our DSD approach introduces no sampling bias in the estimated coefficients, nor prevents conclusions on the causal effects of the treatment from being drawn, even though it does not use random sampling. This may sound surprising to some, given the prevalence of random sampling in clinical trials. However, keep in mind that fundamental DOE principles involve a very similar "nonrandom" mechanism. In any DOE, one judiciously selects a specific combination of factor settings for each experimental run, and these factor settings are not chosen randomly. They are, however, chosen without knowledge of the yet-to-be-realized response value for that run, and this is what avoids the introduction of any bias in the estimated factor effects due to this type of non-random sampling. Otherwise, if DOE principles introduced a bias, then DOE would not be used as ubiquitously as it is. Something quite similar occurs in our DSD approach. We judiciously select a specific combination of patient covariate data (analogous to the factor settings) for each selected patient (analogous to each experimental run) in the entire sample (analogous to the entire experiment), and the patients are selected without consideration of their yet-to-be-realized response data. Consequently, like in DOE, the nonrandom designed sample selection in our DSD approach does not cause any biases. In fact, the judicious (nonrandom) sample selection is precisely what results in a more informative data set being collected than would random sampling, much like how using optimal DOE principles results in a much more informative experiment than if the factor settings were chosen randomly.

We close with some words on the robustness of our approach to missing data and selected patients who opt not to participate in the study. Because we are selecting patients from a large database with patient covariate values, for any patient with missing covariate data we can simply exclude them from consideration. This ensures that all of the selected patients have no missing covariate data. Regarding patients who are selected for either the treatment or control groups but opt not to participate, this clearly affects the information content in the final (participating) study cohort and nullifies its optimality. By design of the DSD approach (because after selecting the sample of 2*n* patients we nearly randomly allocate them to treatment and control groups), we expect that the difference in participation rates between patients selected for the treatment and control groups should be relatively tractable. As future work, we are currently investigating how to incorporate this information into the DSD design strategy so the final participating sample has near-optimality properties.

**Acknowledgement**

This work was supported by the National Science Foundation under Grant #CMMI-1436574, and institutional funding at Northwestern University through Center for Engineering and Health, and Department of Industrial Engineering and Management Science.

**References**

1  Schulz KF, Grimes DA. Generation of allocation sequences in randomised trials: chance, not choice. *The Lancet.* 2002;359(9305):515-519.
2  Lachin JM, Matts JP, Wei L. Randomization in clinical trials: conclusions and recommendations. *Control Clinical Trials.* 1988;9(4):365-374.
3  Zhang Y, Rosenberger WF, Smythe TR. Sequential monitoring on randomization tests: stratified randomization. *Biometrics.* 2007;63(3):865-872.
4  Hu F, Rosenberger WF. *The Theory of Response-Adaptive Randomization in Clinical Trials.* Vol. 525, New Jersey: John Wiley & Sons; 2006.
5  D'Agostino RB. Tutorial in biostatistics: propensity score methods for bias reduction in the comparison of a treatment to a non-randomized control group. *Statistics in Medicine.*1998; 17: 2265-2281.
6  Begg CB, Iglewicz B. A treatment allocation procedure for sequential clinical trials. *Biometrics.* 1980;36(1):81-90.
7  Atkinson A. Optimum biased coin designs for sequential clinical trials with prognostic factors. *Biometrika.* 1982;69(1):61-67.
8  Begg CB, Kalish LA. Treatment allocation for nonlinear models in clinical trials: the logistic model. *Biometrics.* 1984;40(2):409-420.

9   Atkinson A, Biswas A. Adaptive biased-coin designs for skewing the allocation proportion in clinical trials with normal responses. *Statistics in Medicine.* 2005;24:2477-2492.

10  Antognini AB, Zagoraiou M. The covariate-adaptive biased coin design for balancing clinical trials in the presence of prognostic factors. *Biometrika.* 2011;98(3):519-535.

11  Montgomery D. *Design and Analysis of Experiments.* 8th ed. New York: John Wiley & Sons; 2012.

12  Pukelsheim F. *Optimal Design of Experiments (Classics in Applied Mathematics).* Vol. 50. Sociery for Industrial and Applied Mathematic; 2006.

13  Fedorov V. *Theory of Optimal Experiments (Probability and mathematical statistics).* New York: Academic; 1972.

14  Goos P, Jones B. *Optimal Design of Experiments: A Case Study Approach.* Wiley; 2011.

15  Meyer RK, Nachtsheim CJ. The coordinate-exchange algorithm for constructing exact optimal experimental designs. *Technometrics*. 1995;37:60–69.

16  Ouyang L, Apley D, Mehrotra S. A design of experiments approach to validation sampling for logistic regression modeling with error-prone medical records. *Journal of American Medical Informatics Association*, in press, 2015.

17  Rao CR. *Linear Statistical Inference and Its Applications.* 2$^{nd}$ ed. New York: John Wiley & Sons; 1973.

18  Frazer RA, Duncan WJ, Collar AR. *Elementary Matrices, and some Applications to Dynamics and Differential Equations.* Cambridge Univ. Press, 7th (paperback) printing; 1963.

19  Zhang FZ. *The Schur Complement and Its Applications.* Springer; 2005.

20  Harville DA. *Matrix Algebra from a Statistician's Perspective.* New York: Springer; 1997.

21  Sherman J, Morrison WJ. Adjustment of an inverse matrix corresponding to a change in one element of a given matrix. *The Annals of Mathematical Statistics.*  1950; 21:124–127.

Table 1. **Performance comparison of our DSD approach versus random sampling. The numbers in the table are MSE values for the estimated parameters for the treatment factor and its interactions with the covariates, averaged across 10,000 MC replicates.**

|  | $\hat{\alpha}$ | $\hat{\beta}_1$ | $\hat{\beta}_2$ | $\hat{\beta}_3$ | $\hat{\beta}_4$ |
| --- | --- | --- | --- | --- | --- |
| **random sampling** | 9.07E-05 | 9.33E-05 | 9.63E-05 | 9.35E-05 | 9.96E-05 |
| **DSD** | 9.71E-05 | 3.85E-05 | 2.71E-05 | 2.91E-05 | 2.05E-05 |